\documentclass[prd,showpacs,preprintnumbers,floats,floatfix,nofootinbib]{revtex4-1}
\usepackage{graphicx}
\usepackage{dcolumn}
\usepackage{amsmath,amsthm,amssymb}
\usepackage{subfigure}
\usepackage{color}

\begin{document}

\title{The affine-null formulation
of the gravitational equations: spherical case}

\author{ J. A. Crespo${}^{1}$, H.~P. de Oliveira${}^{1}$, J. Winicour${}^{2,3}$
}
\affiliation{
${}^{1}$ Departamento de F{\'{\i}}sica Te\'orica, Instituto de F{\'{\i}}sica\\ CEP 20550-013. Rio de Janeiro, RJ, Brazil\\
${}^{2}$ Department of Physics and Astronomy \\
University of Pittsburgh, Pittsburgh, PA 15260, USA\\
${}^{3}$ Max-Planck-Institut f\" ur
Gravitationsphysik, Albert-Einstein-Institut, \\
14476 Golm, Germany \\
}  

\begin{abstract}

A new evolution algorithm for the characteristic 
initial value problem based upon an affine parameter rather than the
areal radial coordinate used in the Bondi-Sachs formulation is applied
in the spherically symmetric case to
the gravitational collapse of a massless scalar field. The advantages over
the Bondi-Sachs version are discussed, with particular emphasis on the
application to critical collapse.
Unexpected quadratures lead to a simple evolution algorithm
based upon ordinary differential equations which can be integrated along the
null rays. For collapse to a black hole in a Penrose compactified spacetime, these equations are
regularized throughout the exterior and interior of the horizon up to the final singularity.
They are implemented as a global numerical 
evolution code based
upon the Galerkin method. New results regarding
the global properties of critical collapse are presented.

\end{abstract}


\maketitle

\section{Introduction}

The Bondi-Sachs formulation
of Einstein equations~\cite{bondi,sachs},
in which the coordinates are adapted to the null geodesics of
the spacetime, provided historic
and convincing evidence that the emission of gravitational waves is accompanied by mass
loss from the system. (For a review see~\cite{bs_scolar}.)
A technical limitation in the application of the
Bondi-Sachs formulation arises from the use of an areal radial coordinate
to parameterize the outgoing null geodesics. The areal
coordinate becomes singular on and inside
the event horizon so that the Bondi-Sachs formalism is
only applicable in the exterior of the horizon. An alternative approach~\cite{affine}
replaces the areal coordinate by
an affine parameter. 

The difference in behavior between an areal coordinate $r$ and an affine parameter $\lambda$ 
arises from focusing effects on the null rays. The affine coordinate $\lambda$
only becomes singular at points where the null rays
intersect, e.g. caustics, whereas the areal coordinate $r$ also becomes
singular at points where the expansion of the null rays vanish, i.e. where $\partial_\lambda r=0$. 
The Bondi-Sachs formulation was originally adopted for  developing the  PITT null
code~\cite{isaac,highp} for simulating gravitational wave
production because the hierarchical structure
of its system of equations allows them
to be integrated sequentially for one variable at a time along the
outgoing null geodesics.
This hierarchical structure greatly simplifies the evolution algorithm
and is thought to underly its stability.
An affine parameter coordinate was not
adopted because this hierarchical integration
structure was apparently broken. However, by
introducing a (not so obvious) choice of variables,
it was recently shown
how the hierarchical structure of the affine-null system can be regained~\cite{affine}. 

An early triumph of numerical relativity was Choptuik's discovery
of critical phenomena in the spherically symmetric gravitational
collapse of a massless scalar
wave~\cite{choptuik}.
Critical collapse marks the threshold between a system collapsing
to form a black hole or expanding to form an asymptotically
Minkowskian state. 
The use of an areal coordinate for studying critical collapse is an impediment because of its
singular nature at the event horizon.
In the Bondi-Sachs treatment of spherically symmetric gravitational collapse this is not so serious
because the event horizon forms at a single
retarded time, i.e. simultaneously in all 
radial null directions from the center of symmetry. However, the ability to
penetrate the event horizon, as by the affine-null system, is critical in attacking
the non-symmetric case, where
the event horizon forms at different retarded
times for different angles.
Some initial results have been obtained for the critical
collapse of  axisymmetric gravitational
waves~\cite{abev1,abev2} but fundamental
questions remain that have relevance for cosmic censorship
and quantum gravity.
See~\cite{gundlach1,gundlach2} for reviews and discussions of how the critical solution
acts as an attractor for this problem. The affine-null system offers
promising potential for attacking this problem. As a first attempt
in this direction we apply it here to the spherically
symmetric Choptuik problem. The affine-null system has
also been recently applied to the double null characteristic
initial value problem~\cite{maedler}.

Most numerical studies of critical collapse have used
Cauchy evolution
codes, with collapse to a black hole monitored by
the formation of an apparent horizon. A notable
exception by P{\" u}rrer et al.~\cite{purrer} treats
the spherically symmetric Choptuik problem by
means of the Bondi-Sachs formulation. 
Their use of a compactified grid including future
null infinity ${\mathcal I}^+$ allowed them to study how the Bondi
mass and other features of asymptotic flatness
behave on approach to the critical solution, which is
not asymptotically flat. Using mesh refinement, they
confirmed the formation of the
discrete self similarity (DSS) on approach to the critical state and
resolved the fine structure in the universal scaling law for the mass.
Their investigation using the Bondi-Sachs
formalism was limited to the spacetime exterior
to black hole formation.
Other studies of the Choptuik problem using
characteristic evolution codes based upon
double null coordinates were confined to
portions of the spacetime which did not include a
compactification of ${\mathcal I}^+$~\cite{stewart,garfinkle,frolov}.
The afffine-null approach which we  utilize here is applicable to
the entire  exterior
spacetime extending to (compactified) ${\mathcal I}^+$
and to the interior of the black hole extending
to the final singularity.

For the spherical symmetric Choptuik
problem with gravity coupled to a massless field $\Phi$,
the affine-null system has intriguing features. The procedure
for organizing the evolution system
into hierarchical form allows two unexpected quadratures
which lead to a simple evolution algorithm based
upon ordinary differential equations
which can be integrated along the outgoing null rays. The details of an
evolution algorithm valid in the exterior of the event horizon are described in
Sec.~\ref{sec:equations}. Then, in Sec.~\ref{sec:inside},
we introduce renormalized variables which lead to a
well-behaved evolution algorithm
for the entire spacetime
extending from the initial null hypersurface to
the final singularity.

Our study involves unresolved global
aspects of the Choptuik problem, such as the question
whether there is a Bondi mass gap and the effect of
a non-zero Newman-Penrose constant~\cite{npc}
for the scalar field.
The theory underlying these issues is
discussed in Sec.~\ref{sec:physical}. 

The affine-null evolution algorithm is
implemented as a numerical 
evolution code based
upon the Galerkin method. Domain decomposition techniques
are developed to enhance resolution. The numerical
methods are described in Sec.~\ref{sec:galerk}.
New numerical results regarding
critical collapse are presented in Sec.~\ref{sec:numerical}

\section{Spherical symmetry: basic equations}
\label{sec:equations}

The affine-null system~\cite{affine} is based upon a family
of outgoing null hypersurfaces $u=const$
with angular coordinates $x^A$ labeling the null rays and an affine parameter
$\lambda$ to coordinatize points along the rays. 
Here we consider the spherically symmetric case based upon the null hypersurfaces
emanating from the central worldline ${\cal W}$, with
regularity conditions at the vertex. 
In affine-null spherical coordinates
$x^a=(u,\lambda,x^A)$, $x^A=(\theta,\phi)$, the line element takes the form 
\begin{equation}
   g_{ab}dx^a dx^b= -\mathcal{V} du^2 - 2 du d\lambda + r^2 (d\theta^2+\sin^2\theta d\phi^2) 
    \label{eq:naffmet}
\end{equation}
where the metric functions $(\mathcal{V},r)$ depend
upon $u$ and $\lambda$ and $r$ is the areal radius of the
null cones. 
Here the affine freedom $\lambda\rightarrow A(u,x^A)\lambda +B(u,x^A)$ has been used
to prescribe the normalization  $(\nabla^a u)\nabla_a  \lambda =-1$ and to set $\lambda= 0$
on ${\cal W}$.

In order to investigate event horizon formation, we introduce coupling to a massless scalar
field in the form
\begin{equation}
              R_{ab}=\Phi_{,a} \Phi_{,b} \, ,
\end{equation}
where $R_{ab}$ is the Ricci tensor. (We use the shorthand notation
$\partial_a F = F_{,a}$ for derivatives.)

In spherical symmetry,
a complete system of equations for the
affine-null system~\cite{affine} then reduces to
\begin{equation}
               r_{,\lambda\lambda}
              =-\frac{r}{2}\Phi_{,\lambda} \Phi_{,\lambda} \, ,
              \label{eq:r}
\end{equation}
\begin{equation}
   (  \mathcal{V}rr_{,\lambda})_{,\lambda}= 1
      + 2( r^2)_{,\lambda u}
        \label{eq:V} 
\end{equation}
and the scalar wave equation $\Box_g \Phi=0$, which takes the form
\begin{equation}
   (r^2 \Phi_{,u})_{,\lambda} + (r^2 \Phi_{,\lambda})_{,u}
     - (r^2{\cal V}\Phi_{,\lambda})_{,\lambda} =0 \, .
     \label{eq:wave}
\end{equation}
Following the procedure in~\cite{affine},
we introduce the variables
\begin{eqnarray}
 \rho &\equiv& r_{,u} \\
 \mathcal{Y} &\equiv& \mathcal{V} - 2 \frac{\rho}{r_{,\lambda}}  
 \label{eq:Y}\\
 K &\equiv& 2(r_{,\lambda} \Phi_{,u} - \rho \Phi_{,\lambda}) .
\end{eqnarray}
Then (\ref{eq:r})-- (\ref{eq:wave}) take the hierarchical form
\begin{equation}
               r_{,\lambda\lambda}
              =-\frac{r}{2}\Phi_{,\lambda} \Phi_{,\lambda}
              \label{eq:sr}
\end{equation}
\begin{equation}
        \bigg (  {\cal Y}(r^2)_{,\lambda} \bigg)_{,\lambda}= 2 
        \label{eq:sY}
\end{equation}
\begin{equation}
       r\bigg( \frac{rK}{r_{,\lambda}} \bigg )_{,\lambda} 
        -(r^2 {\cal Y} \Phi_{,\lambda})_{,\lambda} =0.
       \label{eq:sK}
\end{equation}
Given $\Phi(u,\lambda)$, these equations can be integrated sequentially to determine
$r$, ${\cal Y}$ and $K$, in that order.

An evolution algorithm valid in the exterior of the horizon
can be formulated by taking
the $u$-derivative of (\ref{eq:sr}),
\begin{equation}
      \rho_{,\lambda\lambda}= -\frac{1}{2} \rho \Phi_{,\lambda} \Phi_{,\lambda}
      - r \Phi_{,\lambda} \Phi_{, u\lambda} ,
         \label{eq:srho}
\end{equation}
which can be expressed in the form
\begin{equation}
    \bigg ( \frac{ \rho}{r_{,\lambda} }\bigg )_{,\lambda\lambda}=
    -\frac{r\Phi_{,\lambda}}{2r_{,\lambda}}
         (\frac{K}{r_{,\lambda}})_{,\lambda}.
      \label{eq:serho}
\end{equation}

Equations  (\ref{eq:sr})--(\ref{eq:sK}) and (\ref{eq:serho}) are regular in the exterior
of the event horizon where $r_{,\lambda}>0$.
They gives rise to the following evolution algorithm. Specify the initial data 
\begin{equation}
               \Phi(0,\lambda) \, ,  \quad u=0,  \,  \lambda \ge 0 
\end{equation}
and impose the regularity conditions at the central geodesic
\begin{equation}
   r=0, \quad  r_{,\lambda} =1,  \quad  {\cal Y}=1,
    \quad rK=\rho= \rho_{,\lambda}=0, 
    \quad u\ge 0, \,  \lambda=0.
    \label{eq:vertex}
\end{equation}
Using this initial data, integrate (\ref{eq:sr})--(\ref{eq:sK}) and (\ref{eq:serho}) in sequential
order to determine the initial values of the variables $(r,{\cal Y}, K ,\rho)$.
Given this initialization and the vertex regularity conditions, the system can then be evolved
to $u=\Delta u$ by a finite difference approximation to determine
$r(\Delta u,\lambda)$ algebraically from $\rho$ and to determine
$\Phi(\Delta u,\lambda)$ algebraically
from $K$. Then (\ref{eq:sY}), (\ref{eq:sK})
and (\ref{eq:serho}) can be integrated in sequential order to determine ${\cal Y}$, $K$ and
$\rho$  at $u=\Delta u$. Repetition of this process provides the
evolution algorithm.

In fact, this procedure can be simplified since
(\ref{eq:sY}) integrates to give
\begin{equation}
   {\cal Y}= \frac{\lambda}{r r_{,\lambda}}
   \label{eq:Y2}
\end{equation}
so that ${\cal Y}$ can be eliminated 
and (\ref{eq:sK}) reduces to
\begin{equation}      
    \bigg( \frac{rK}{r_{,\lambda}} \bigg )_{,\lambda} 
 =\frac{1}{r}\bigg(\frac{\lambda r\Phi_{,\lambda}}
      {r_{,\lambda}}
    \bigg )_{,\lambda} .
       \label{eq:rsK}
\end{equation}

In the exterior of the event horizon, we assume that the initial data
$\Phi(u_0,\lambda)$ has an asymptotic $1/\lambda$ expansion
$\Phi(u_0,\lambda)=\alpha(u_0)\lambda^{-1} +\beta(u_0)\lambda^{-2}+ \dots$
so that it is consistent with asymptotic flatness. 
This asymptotic behavior is preserved in the
exterior spacetime by the evolution equations, i.e.
\begin{equation}
  \Phi(u,\lambda)= \alpha(u)\lambda^{-1} +\beta(u)\lambda^{-2}+O(\lambda^{-3}) .
  \label{eq:phiexp}
\end{equation}
Integration of (\ref{eq:sr}) then leads to the asymptotic expansion
\begin{equation}
     r = H(u)\lambda + R(u) - \frac{H\alpha^2}{4\lambda}+ O(\lambda^{-2}) .
     \label{eq:rexp}
\end{equation}

\section{Regularized evolution inside the horizon}
\label{sec:inside}

The expansion of the outgoing null cones is
$\Theta^+=2r_{,\lambda}/r$.
The regularity conditions at the vertex
(\ref{eq:vertex}) require $r|_{\lambda=0}=0$ and
$r_{,\lambda}|_{\lambda=0}=1$. Integration of  (\ref{eq:sr}) then implies
$r_{,\lambda}$ decreases monotonically with $\lambda$, with the behavior
near the vertex
\begin{equation}
  r=\lambda -\frac{\lambda^3}{12}
  \big(\Phi_{,\lambda}(u,0)\big)^2 +O(\lambda^4).
\end{equation}
The radially inward pointing null vector is 
\begin{equation}
n^a\partial_a= \partial_u
-\frac{{\cal V}}{2} \partial_\lambda
= \partial_u-\frac{(\lambda+2r\rho)}{2rr_{,\lambda}}\partial_\lambda,
\end{equation}
normalized by $n^a \partial_a u=1$. The expansion of the ingoing null cones is
\begin{equation}
  \Theta^-=\frac{4}{r} n^a \partial_a r
      = -\frac {2\lambda} {r^2},
\end{equation}
which is everywhere negative.

In the exterior untrapped region
where $\Theta^+ > 0$,
$H(u)=r_{,\lambda}\big |_{\lambda=\infty}$
satisfies $1\ge H(u) > 0$,
with equality $H=1$
only in the trivial case $\Phi(u,r) =const$. In the supercritical case,
on approach to the event horizon at $u=u_E$,
\begin{equation}
   \lim_{u\rightarrow u_E} H(u) =0.
\end{equation}
Inside the event horizon there is an apparent horizon traced out by
$\lambda=\lambda_A(u)$, where $r_{,\lambda}(u,\lambda_A) =0$. After formation of the
apparent horizon, the outgoing null cones from the central worldline
recollapse to a singularity at $r=0$ at a finite value of $\lambda$.
Consequently, terms in the evolution equations (\ref{eq:serho}) and
(\ref{eq:rsK}) containing $1/r_{,\lambda}$ become singular at the apparent
horizon. This is not a true singularity and there is
a way to regularize the evolution system.

In order to regularize (\ref{eq:rsK}), we set
\begin{equation}
  L  =  \frac{rK-\lambda \Phi_{,\lambda}}{r_{,\lambda}} 
 = 2r\Phi_{,u} - \frac{(2r\rho +\lambda)\Phi_{,\lambda}}{r_{,\lambda}} .
  \label{eq:hL}
\end{equation}
As a result, (\ref{eq:rsK}) becomes  
\begin{equation}
     L_{,\lambda}  = \frac{\lambda \Phi_{,\lambda}}{r} \, .
              \label{eq:Lev}
\end{equation} 
The right hand side of (\ref{eq:Lev}) is regular everywhere in the exterior spacetime,
including ${\mathcal I}^+$, and everywhere regular
inside the event horizon up to the final singularity.
Thus, as a result of the integration of (\ref{eq:Lev}),
$L$ is also regular everywhere.

In order to regularize the $\rho$ equation (\ref{eq:serho}) we
introduce the variable
\begin{equation}
      P = \frac{2r\rho +\lambda}{rr_{,\lambda}} \, .
\end{equation}
Then, after considerable algebra involving the use of (\ref{eq:Lev}), we rewrite (\ref{eq:serho}) as
\begin{equation}
   P_{,\lambda\lambda} 
   =  \frac{2\lambda r_{,\lambda}}{r^3} - \frac{2 }{r^2}
       + \frac{( L^2)_{,\lambda} }{2\lambda} .        
             \label{eq:2hPev}
\end{equation}
The validity of (\ref{eq:2hPev}) can be checked by straightforward calculation.

The right hand side of (\ref{eq:2hPev}) is regular everywhere in
the spacetime up to the final singularity.
Thus as a result of the integration of (\ref{eq:2hPev}),
using the vertex regularity conditions (\ref{eq:vertex}),
$P$ is also regular throughout the spacetime.

It is remarkable that (\ref{eq:2hPev}) has a first integral. Multiplication
by $\lambda$
and use of the vertex regularity conditions, which
require $P(u,0) =1$ and  $ L(u,0) =0$,
leads after integration to
\begin{equation}
 \left(\frac {P}{\lambda}\right)_{,\lambda} = 
 -\frac{1}{r^2} +\frac{ L^2}{2\lambda^2} \, .
                \label{eq:hPev}
\end{equation}
Since  $P/\lambda$ is singular at the vertex, we introduce the
variable
\begin{equation}
Q = (P-1)/\lambda = \frac{2r\rho +\lambda-rr_{,\lambda}}{\lambda rr_{,\lambda}} \, .
\label{eq:cp}
\end{equation}
Then (\ref{eq:hPev}) becomes
\begin{equation}
 Q_{,\lambda} = \frac{1}{\lambda^2} -\frac{1}{r^2} + \frac{ L^2}{2\lambda^2}.
                \label{eq:cPev}
\end{equation}

In summary, (\ref{eq:r}),
 (\ref{eq:Lev}) and (\ref{eq:cPev}) lead to the evolution system consisting of the three
 hypersurface equations
\begin{equation}
r_{,\lambda\lambda} =
- \frac{r}{2}\left(\Phi_{,\lambda}\right)^2 
         \label{eq:r2}
\end{equation} 
\begin{equation}
     L_{,\lambda} = \frac{\lambda \Phi_{,\lambda}}{r}
              \label{eq:sLev}
\end{equation} 
\begin{equation}
   Q_{,\lambda} = \frac{1}{\lambda^2} -\frac{1}{r^2} +\frac{ L^2}{2\lambda^2} ,
                \label{eq:xcPev}
\end{equation}
where the definitions of $L$ and $Q$, (\ref{eq:hL}) and (\ref{eq:cp}), combine to give the
evolution equation
\begin{equation}
\Phi_{,u} = \frac{\lambda Q}{2}\Phi_{,\lambda} + \frac{1}{2}\Phi_{,\lambda} +\frac{L}{2 r} . 
\label{eq:KG}
\end{equation}

The ordinary differential equations (\ref{eq:r2})--(\ref{eq:xcPev}) can be integrated along the outgoing null rays.
The resulting solution is regular at the vertex, where $Q(u,0)=0$ and $ L(u,0)=0$, is regular at ${\mathcal I}^+$
and is regular inside the horizon up to the final singularity. The system gives rise to the following evolution algorithm
which  covers the entire spacetime to the future of the initial null hypersurface. Given the initial data $\Phi(u_0,\lambda)$,
integrate (\ref{eq:r2}) to determine the initial value $r(u_0,\lambda)$. Then (\ref{eq:sLev}) and (\ref{eq:xcPev}) can be
integrated in sequential order to determine $L(u_0,\lambda)$, and $Q(u_0,\lambda)$. With these values
$\Phi_{,u}(u_0,\lambda)$ is readily obtained from (\ref{eq:KG}). Thus a finite difference approximation determines
$\Phi(u_0+\Delta u,\lambda)$. The repetition of this process provides a global evolution algorithm, whose numerical
implementation is described in Sec.~\ref{sec:galerk}.

Note that in the linearized limit, i.e. up to terms linear in
$\Phi$, (\ref{eq:sLev}) reduces to the flat space, spherically symmetric
scalar wave equation and (\ref{eq:xcPev}) implies $Q=0$. 

\section{Physical properties}
\label{sec:physical}

Bondi time $u_B$, i.e. the time intrinsic to
an inertial observer at null infinity,
is related to the central proper time by
$\partial  u_B/\partial u=1/H$,
where $H(u)=r_{,\lambda}(u,\lambda)|_{\lambda=\infty}$.
In the supercritical case of event horizon formation at
$u=u_E$, $H(u)\rightarrow 0$ as $u\rightarrow u_E$.
Thus, although the horizon forms at a finite central
time it forms at an infinite Bondi time, $u_B \rightarrow \infty$, in accord with the
infinite redshift of a distant observer.

\subsection{No scalar hair}

The regularity of the affine-null evolution system implies that quantities
that have finite $u$-derivatives on the event horizon, e.g  $ \Phi_{,u}(u_E,\lambda)$,
must have vanishing Bondi-time derivative so that
$\partial_{u_B} \Phi(u_B,\lambda)\rightarrow 0$ as
$u_B \rightarrow \infty$. 
This is consistent with the results of Christodoulou~\cite{christo} obtained by
applying the methods of analysis to the Bondi formulation of the Einstein-scalar
equations. The scalar monopole moment
is defined by
\begin{equation}
   A(u):=\lim_{r \rightarrow \infty} r \Phi 
\end{equation}
so, referring to the asymptotic expansions (\ref{eq:phiexp}) of $\Phi$ and (\ref{eq:rexp}) of $r$,
\begin{equation}
    A(u) = \lim_{\lambda \rightarrow \infty} H \lambda \Phi 
     =H(u) \alpha(u) .
\end{equation}
Of special importance, since $H(u_E)=0$,
it follows that $A(u) \rightarrow 0$
as $u\rightarrow u_E$, i.e. as
$u_B \rightarrow \infty$, in accordance with the
``no hair'' property of the black hole.  

\subsection{Newman-Penrose constant}

In Bondi coordinates, $\Phi$ has the asymptotic
expansion 
$$ \Phi = \frac {A}{r} + \frac {c_{NP}}{r^2} + O(r^{-2})
$$
where $c_{NP}$ is the Newman-Penrose constant~\cite{npc} for the scalar field. In order to express
$c_{NP}$ in affine-null coordinates we write 
$$ c_{NP}=-r^2 \partial_r  (r\Phi) |_{r=\infty}
 = -\frac {r^2}{r_{,\lambda}}  \partial_\lambda  (r\Phi) |_{r=\infty}.
$$
Then, from the asymptotic expansions (\ref{eq:phiexp}) of $\Phi$ and (\ref{eq:rexp}) of $r$,
we obtain
\begin{equation}
c_{NP} = H({\beta H+\alpha R}). \label{eq:NP}
\end{equation}

In order to verify that $\partial_u c_{NP} =0$ we consider
the asymptotic expansion of the evolution 
equation (\ref{eq:sLev}) for $L$. From (\ref{eq:hL}),
we have
\begin{equation}
  L = 2(H\alpha)_{,u}
  + \frac{1}{H\lambda} \big( 2(RH\alpha+H^2 \beta)_{,u}
  +\alpha \big ) + \dots \, .
  \label{eq:ahatl}
\end{equation}  
But (\ref{eq:sLev}) implies
\begin{equation}
    L_{,\lambda} =-\frac{\alpha}{H\lambda^2} 
      +\big ( \frac{R\alpha}{H} 
      -2\beta\big ) \frac{1}{H\lambda^3}+\dots \, .
      \label{eq:Llamx}
\end{equation}
Comparison of (\ref{eq:ahatl}) with (\ref{eq:Llamx}) gives
$(RH\alpha+H^2 \beta)_{,u} =0$, in agreement with
the conservation law $\partial_u c_{NP} =0$.

As a result, the expansion (\ref{eq:ahatl}) reduces to 
\begin{equation}
   L = 2(H\alpha)_{,u}
  + \frac{ \alpha }{H\lambda}  + \dots ,
  \label{eq:cahatl}
\end{equation} 
and, to the next order,
\begin{equation}
    L =2(H\alpha)_{,u}+\frac{\alpha}{H\lambda} 
    -\big ( \frac{R\alpha}{H} 
      -2\beta\big ) \frac{1}{2H\lambda^2}+\dots \, .
\end{equation}
 
In the numerical simulations we consider
initial data of the form
\begin{equation}
\Phi(0,\lambda) =  \frac{\epsilon}{a^2+\lambda^2},
\label{eq:npdata}
\end{equation}
for which $\alpha|_ {u=0}=0$ and  
$c_{NP}=\epsilon H^2 |_ {u=0}$ is a non-zero Newman-Penrose constant.
This evolves to form a black hole
for sufficiently large $\epsilon$. The numerical behavior
of the Newman-Penrose constant for slightly subcritical
and slightly supercritical initial data is plotted
in Sec.~\ref{sec:numerical}.
 
It might at first seem paradoxical that the Newman-Penrose
constant must be conserved in the subcritical case where
the scalar field vanishes, i.e.
$\Phi\rightarrow 0$,  as $u\rightarrow \infty$. This is
explained by the non-uniform behavior of the limits
$u\rightarrow \infty$ and $\lambda\rightarrow \infty$,
which cannot be interchanged. As an example,
consider Minkowski space where $\lambda =r$
and non-singular solutions of the wave equation take
the form
$$ \Phi =\frac{f(t+r) -f(t-r)}{r} 
= \frac{f(u+2r) -f(u)}{r}  ,
$$
where $f$ is a smooth function.
Then the initial data (\ref{eq:npdata}) correspond
to the flat space solution
$$ \Phi = \frac{2\epsilon}{r} \bigg( \frac{u+2r}
 {a^2 +(u+2r)^2}  -\frac{u} {a^2 +u^2} \bigg )
 $$
 with limit
 $$\lim_{u\rightarrow \infty} \Phi =0,
 $$
 so that the scalar field decays to zero but
 $$c_{NP}=-\lim_{r\rightarrow \infty} r^2 \partial_r  (r\Phi) = \epsilon,
 $$
 independent of $u$. 

\subsection{The Bondi mass}

In spherical symmetry the Misner-Sharp
mass function $m(u,\lambda)$ is defined
as~\cite{misnersharp}
\begin{equation}
1 - \frac{2 m}{r} = g^{\alpha\beta}r_{,\alpha}r_{,\beta} \, .
\end{equation}
In affine-null coordinates
$g^{\alpha\beta} r_{,\alpha}r_{,\beta}= (\mathcal{V}r_{,\lambda} 
- 2 \rho) r_{,\lambda}=\mathcal{Y}(r_{,\lambda})^2 =\lambda r_{,\lambda}/r$,
where we have used (\ref{eq:Y}) and(\ref{eq:Y2}). Therefore,
\begin{equation}
 m(u,\lambda) =\frac{1}{2}(r - \lambda r_{,\lambda}).
\end{equation}
The Bondi mass of the system is determined by taking the asymptotic limit\
\begin{equation}
M_B (u) = \lim_{\lambda \rightarrow \infty} m(u,\lambda) = \frac{R(u)}{2} \, ,
\end{equation}
where $R(u)$ is obtained from the asymptotic
expansion  of $r$ according to (\ref{eq:rexp}).

In order to recover the mass loss equation due
to scalar radiation we consider the asymptotic
behavior of the evolution equation for $Q$.
From the definition (\ref{eq:cp}) of
$Q$ and the asymptotic expansions
(\ref{eq:rexp}) and (\ref{eq:phiexp}),
we obtain
\begin{equation}  
    Q = \frac{2H_{,u}}{H}
   +\frac{1-H^2+2HR_{,u}}{H^2 \lambda} + \dots \, .
\label{eq:ahatp}    
\end{equation}
Using  (\ref{eq:cahatl}),  the evolution equation  (\ref{eq:xcPev}) gives
\begin{equation}
  Q_{,\lambda} =
  \frac{H^2 -1+2[(HH\alpha)_{,u}]^2}{H^2\lambda^2} 
  +\dots \, .
 \end{equation}
Comparison with (\ref{eq:ahatp}) gives the
mass loss equation
\begin{equation}
    M_{B,u} 
    =\frac{1}{2} R_{,u}=- \frac{1}{2}H[(H\alpha)_{,u}]^2 \, .
    \label{eq:mloss}
\end{equation}

In the supercritical case, it follows that
\begin{equation}
   \lim_{u\rightarrow u_E}  M_{B,u}
   =\frac{1}{2}\lim_{u\rightarrow u_E} R_{,u}=0
\end{equation}
on  approach to the event horizon.
The mass loss equation (\ref{eq:mloss}) provides
a convenient test of code accuracy, as  presented
in Sec.~\ref{sec:numerical}.

\subsection{Approach to horizon formation}

Nontrivial data $\Phi(u_0,\lambda)$ implies $0\le H(u_0)<1$.
In general, $H_{,u}$ can either be positive or negative. In the subcritical
case in the limit that the spacetime becomes flat $H(u=\infty,\lambda)=1$,
so that $H_{,u}>0$ at late times.

In the supercritical case, in the limit of horizon
formation
$H(u_E) =0$ so that $H_{,u}(u_E)\le 0$. As a result,
the Bondi mass loss equation (\ref{eq:mloss}) leads to $M_{B,u}(u_E) =0$ and
$$M_{B,uu}(u_E) =-\frac{(H_{,u})^3}{2}\alpha^2 \big |_{u=u_E} \ge  0.
$$
Unless either $H_{,u}(u_E)=0$ or  $\alpha(u_E) = 0$, the inequality
$M_{B,uu}(u_E) > 0$ would hold in the limit
so that $M_B (u_E)$ is a strong minimum.
The numerical results  in  Sec.~\ref{sec:numerical} indicate that  $H_{,u}(u_E)<<0$,
i.e. that $H(u)$ goes to zero at a fast rate.
But, generic numerical results show that
$H_{,u}(u_E)\ne 0$ and 
$\alpha(u_E)\ne 0$, so that, $M_B (u_E)$ is a strong minimum.

Bondi time $u_B$ at null infinity is related to the central proper time by $\partial u_B/\partial u=1/H$.
As a result, since $H_{,u}(u_E)$ is negative, on approach to the event horizon,
$H(u) \sim H_{,u}(u_E) (u-u_E)$ and Bondi time
goes to infinity as 
\begin{equation}
u_B \sim \frac{ \ln(u_E -u)}{ H_{,u}(u_E)}.
\end{equation}

\section{Numerical method: the Galerkin-collocation approach}
\label{sec:galerk}

\begin{figure}[ht]
\begin{center} 
\includegraphics*[scale=0.5]{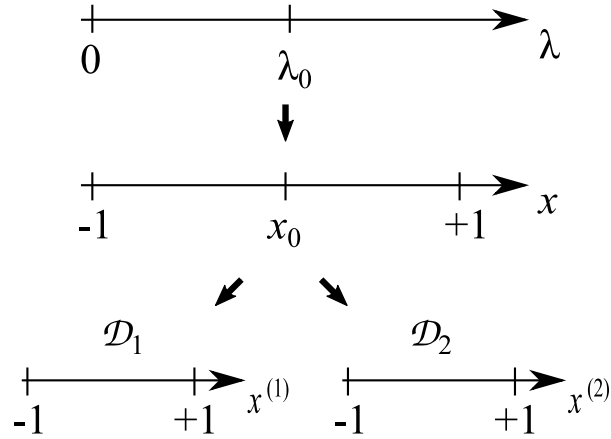} 
\end{center}
\caption{Scheme showing the subdomains $\mathcal{D}_1$, $0 \leq \lambda \leq \lambda_0$,
and $\mathcal{D}_2$, $\lambda_0 \leq \lambda < \infty$. We also present the corresponding
computational domains $-1 \leq x^{(A)} \leq 1$
with $A=1,2$ for each domain. }
\end{figure}

\begin{table*}
\centerline{Table 1}
\medskip

	\centering
		\begin{tabular}{c|c}
		\hline
		\\
		$\mathcal{D}_1: 0 \leq \lambda \leq \lambda_0$ & $\mathcal{D}_2: \lambda_0 \leq \lambda < \infty$\\
		\\
		\hline
		\hline
		\\
		$\Phi^{(1)}(u,\lambda) = \sum\limits_{k=0}^{N_1}\,a^{(1)}_{k}(u) TL^{(1)}_{k}(\lambda)$ & $\Phi^{(2)}(u,\lambda)
		= \sum\limits_{k=0}^{N_2}\,a^{(2)}_{k}(u) \psi^{(2)}_{k}(\lambda)$\\
		\\
$r^{(1)}(u,\lambda) = \lambda + \sum\limits_{k=0}^{N_1}\,b^{(1)}_{k}(u) \lambda^3 TL^{(1)}_{k}(\lambda)$ &
$r^{(2)}(u,\lambda) = \sum\limits_{k=0}^{N_2+2}\,b^{(2)}_{k}(u) \lambda TL^{(2)}_{k}(\lambda)$\\
    \\
$L^{(1)}(u,\lambda) = \sum\limits_{k=0}^{N_1}\,c^{(1)}_{k}(u) \chi^{(1)}_{k}(\lambda)$ & $L^{(2)}(u,\lambda) 
= \sum\limits_{k=0}^{N_2+1}\,c^{(2)}_{k}(u) TL^{(2)}_{k}(\lambda)$\\
		\\
$Q^{(1)}(u,\lambda) = \sum\limits_{k=0}^{N_1}\,f^{(1)}_{k}(u) \chi^{(1)}_{k}(\lambda)$ & $Q^{(2)}(u,\lambda) 
= \sum\limits_{k=0}^{N_2+1}\,f^{(2)}_{k}(u) TL^{(2)}_{k}(\lambda)$\\
\\
		Radial basis function:& Radial basis function:\\
	\\
	$\chi_k^{(1)}(\lambda) = \frac{1}{2}(TL_{k+1}^{(1)}(\lambda)+TL_k^{(1)}(\lambda))$ & $\psi_k^{(2)}(\lambda)
	= \frac{1}{2}(TL_{k+1}^{(2)}(\lambda)-TL_k^{(2)}(\lambda))$\\
	\\
	\hline
	\end{tabular}
		\end{table*}

To integrate the field equations, we have implemented a code based on the Galerkin-collocation method \cite{hpo}
using the domain decomposition technique. In general, single domain spectral methods are very accurate, but
if we are to determine the formation of black holes of infinitesimal masses, it is necessary to establish a spectral version
of mesh refinement provided by dividing the spatial domain into several subdomains.  

We have implemented a simple version of the domain decomposition technique dividing the spatial domain
$0 \leq \lambda < \infty$ into two non-interpolating subdomains, $\mathcal{D}_1: 0 \leq \lambda \leq \lambda_0$
and  $\mathcal{D}_2: \lambda_0 \leq \lambda < \infty$, where $\lambda=\lambda_0$ denotes the interface of these subdomains.
The innovative part of implementing the algorithm is the two-step introduction of the correspondent computational,
as indicated in Fig. 1. In the first step, the physical domain is compactified using the
algebraic map \cite{boyd}

\begin{equation}
\lambda = L_0 \frac{(1+x)}{1-x},
\end{equation}

\noindent so that the interval $0 \leq \lambda < \infty$ corresponds to $-1 \leq x \leq 1$, and $L_0$ is the
map parameter. In the second step, the subdomains $-1 \leq x \leq x_0$ and $x_0 \leq x \leq 1$ are mapped
into the subdomains characterized by $-1 \leq x^{(1)} \leq 1$ and $-1 \leq x^{(2)} \leq 1$, respectively, by
linear maps. For simplicity we have set the location of the interface at
$x=x_0=0$ in the intermediate computational domain,
which is equivalent to setting $\lambda_0=L_0$.

We approximate the relevant fields $\Phi, r, L$ and $Q$ as series with respect to appropriate sets of
basis functions. According to the Galerkin method, each element of the basis functions must satisfy the
boundary conditions of each subdomain. The approximations are shown in Table 1. In these expressions $N_A$, $A=1,2$, are the truncation orders at each subdomain that dictate the number of unknown modes $a^{(A)}_k(u),b^{(A)}_k(u),c^{(A)}_k(u),f^{(A)}_k(u)$. The basis functions $TL_k^{(A)}(\lambda)$ are the
rational Chebyshev polynomials defined at each subdomain by

\begin{eqnarray}
TL_k^{(1)}(\lambda) = T_k\left(x^{(1)}=\frac{3\lambda-L_0}{\lambda+L_0}\right)  \\
\nonumber\\
TL_k^{(2)}(\lambda) = T_k\left(x^{(2)}=\frac{\lambda-3L_0}{\lambda+L_0}\right), 
\end{eqnarray}

\noindent where $T_k(x)$ represents the standard Chebyshev polynomials of order $k$. The basis functions
$\chi^{(1)}_k(\lambda)$ and $\psi^{(2)}_k(\lambda)$ are expressed in terms of the rational Chebyshev polynomials
to satisfy the boundary conditions (cf. Table 1).
The domain decomposition method requires
junction or transmission conditions for the relevant fields at the interface $\lambda=\lambda_0$. These conditions
differ for the hypersurface and evolution equations. Starting with the hypersurface equations for $L$ and $P$, we have

{\small
\begin{eqnarray}
L^{(1)}(u,\lambda_0) = L^{(2)}(u,\lambda_0),\; \left(\frac{\partial L^{(1)}}{\partial u}\right)_{\lambda_0} =
  \left(\frac{\partial L^{(2)}}{\partial u}\right)_{\lambda_0} \\
\nonumber \\
Q^{(1)}(u,\lambda_0) = Q^{(2)}(u,\lambda_0),\; \left(\frac{\partial Q^{(1)}}{\partial u}\right)_{\lambda_0}
 = \left(\frac{\partial Q^{(2)}}{\partial u}\right)_{\lambda_0}.
\end{eqnarray}}

\noindent The particular form of the hypersurface equation for the metric function $r(u,\lambda)$ demands the conditions

{\small
\begin{eqnarray}
r^{(1)}(u,\lambda_0) &=& r^{(2)}(u,\lambda_0),\; \left(\frac{\partial r^{(1)}}{\partial u}\right)_{\lambda_0} 
  = \left(\frac{\partial r^{(2)}}{\partial u}\right)_{\lambda_0} \label{eq_junc_r1}\\
\nonumber \\
\left(\frac{\partial^2 r^{(1)}}{\partial u^2}\right)_{\lambda_0} &=& \left(\frac{\partial^2 r^{(2)}}{\partial u^2}\right)_{\lambda_0}.
\label{eq_junc_r2}
\end{eqnarray}}

For the scalar field, it is necessary to guarantee its continuity at the interface,

\begin{eqnarray}
\Phi^{(1)}(u,\lambda_0) = \Phi^{(2)}(u,\lambda_0).
\end{eqnarray}

\noindent Following Canuto et al. \cite{canuto}, we have adopted the average procedure where both subdomains have the
same weight in the update equation for the interface point. This interface condition is 

{\small
\begin{eqnarray}
(\Phi_{,u})_{\lambda_0} - && \frac{1}{2}\left[\frac{L^{(1)}}{2 r^{(1)}}
+\frac{1}{2}(1+\lambda Q^{(1)})\Phi^{(1)}_{,\lambda}\right]_{\lambda_0} 
- \frac{1}{2}\left[\frac{L^{(2)}}{2 r^{(2)}}
+\frac{1}{2}(1+\lambda Q^{(2)})\Phi^{(2)}_{,\lambda}\right]_{\lambda_0} = 0 .\nonumber \\ 
\end{eqnarray}
}

The final step in establishing the algorithm is to substitute the approximations of Table 1 into the field equations
to form the residual equations in each subdomain. We have followed the collocation method by imposing
that the residual equations vanish at the $N_1$ and $N_2+1$ interior collocation points in the subdomains
$\mathcal{D}_1$ and $\mathcal{D}_2$, respectively. Therefore, there are $N_1+N_2+1$ equations that together
with the transmission conditions provide the same number of unknown coefficients. For the sake of illustration,
consider the  residual equation associated to $r(u,\lambda)$ at the second domain,

\begin{eqnarray}
\mathrm{Res}_{r^{(2)}}(u,\lambda_j) &&= \sum_{k=0}^{N_2+2} b_k^{(2)}\left[(\lambda TL^{(2)}_k)_{,\lambda\lambda}\right]_j
 +\frac{1}{2}r^{(2)}_j \left(\Phi^{(2)}_{,\lambda}\right)_j^2, \nonumber \\
\end{eqnarray}

\noindent for all $j=1,2,..,N_2+1$. Here $r^{(2)}_j$ and $ \left(\Phi^{(2)}_{,\lambda}\right)_j$ are values of these fields at the
collocation points. Thus, we have $N_2+1$ equations and $N_1$ equations from the first and second subdomains, respectively,
which together with three transmission conditions given by (\ref{eq_junc_r1}) and (\ref{eq_junc_r2}) constitute a set of
$N_1+N_2+4$ algebraic equations for an equal number of unknown coefficients $b_k^{(1)}(u)$ and $b_k^{(2)}(u)$.
Repeating a similar procedure for the hypersurface equations (\ref{eq:sLev}) and (\ref{eq:xcPev}), we obtain sets of
algebraic equations for the modes $c_k^{(A)}(u)$ and $f_k^{(A)}(u)$, $A=1,2$. 

Concerning the evolution equation (\ref{eq:KG}),
the vanishing of the corresponding residual equations at the collocation points in both subdomains, together with the
transmission conditions, yield a set of ordinary differential equations for the coefficients $a^{(A)}_k(u)$. 

The hierarchy of the field equations is preserved in the spectral representation. Specifically, once the coefficients
$a^{(A)}_k(u_0)$ are initially fixed, the initial modes $b^{(A)}_k(u_0)$ are determined from the algebraic set described above.
In the sequence, the remaining modes $c^{(A)}_k(u_0)$ and $f^{(A)}_k(u_0)$ can be calculated. Then, the set of ordinary
differential equations determine $a^{(A)}_{k,u}(u_0)$ allowing these modes to be
updated to the next time step. Repetition of this process provides the numerical solution of the field equations.

In order to evolve the self-gravitating scalar we need to specify the initial data
$\Phi_0(\lambda) = \Phi(u=0,\lambda)$
that fix the initial modes $a^{(A)}_k(0)$ in both subdomains. We have chosen the three initial data sets
\begin{equation}
r_0(\lambda)=(1-\epsilon) \lambda + \epsilon \tanh(\lambda) ,
\label{eq:id_1}
\end{equation}
for which $\Phi_0(\lambda)$ is determined from
the hypersurface equation (\ref{eq:r}),
\begin{equation}
  \Phi_0(\lambda)=\frac{\epsilon}{1+\lambda^2} \label{eq:id_2} 
\end{equation}
and
\begin{equation}
  \Phi_0(\lambda)=\frac{\epsilon}{2}\left(TL_{k+1}(\lambda)-TL_{k}(\lambda)\right),  
   \label{eq:id_3} 
\end{equation}
determined from the Chebyshev polynomials by  $TL_{k}(\lambda)=T_k(x=(\lambda-L_0)/(\lambda+L_0))$. Here $\epsilon$
is the parameter that plays the role of the amplitude of the initial scalar field. 

\section{Numerical results}
\label{sec:numerical}

We use the Bondi mass loss equation
(\ref{eq:mloss}) to calibrate the accuracy and convergence of the code for the affine-null system.
Integration of (\ref{eq:mloss})  gives 
\begin{equation}
 M(u) - M_0 = -\frac{1}{2}\int_{u_0}^u\,H \left[(H \alpha)_{,u}\right]^2 du,
\end{equation}
\noindent where $M_0 = M(u_0)$ is the initial Bondi mass and $M_0(u) - M(u)$ is the mass loss evaluated at
retarded time $u$, which equals the energy  radiated in this interval described by the integral.
The numerical test consists in verifying global energy conservation measured by the quantity
$C(u)$~\cite{winicour_92},
\begin{equation}
C(u) \equiv 1-\frac{M(u)}{M_0}-\frac{1}{2M_0}\int_{u_0}^u\,H \left[(H \alpha)_{,u}\right]^2 du,
\end{equation}

\noindent where $\alpha = \lim_{\lambda \rightarrow \infty} (\lambda \Phi)$ and
$H=\lim_{\lambda \rightarrow \infty} r/\lambda$ (cf. (\ref{eq:phiexp}) and (\ref{eq:rexp})).
We measure the numerical deviation from the \textit{exact} result $C(u)=0$
by computing the maximum value of $C(u)$ for evolutions with increasing truncation order.
For this test, we use the
single domain Galerkin-collocation code. We evolve the   initial data (\ref{eq:id_1}) with $\epsilon=0.5$,
corresponding to a subcritical solution, with truncation orders $N=30,40,50,..,90$.  The results are presented in Fig. 2.
For comparison, we have included results of the same test using a similar code based upon the standard
Bondi equations and coordinates. It is clear that the
error for the affine-null scheme decays more rapidly.

\begin{figure}
\includegraphics[width=8cm,height=7cm]{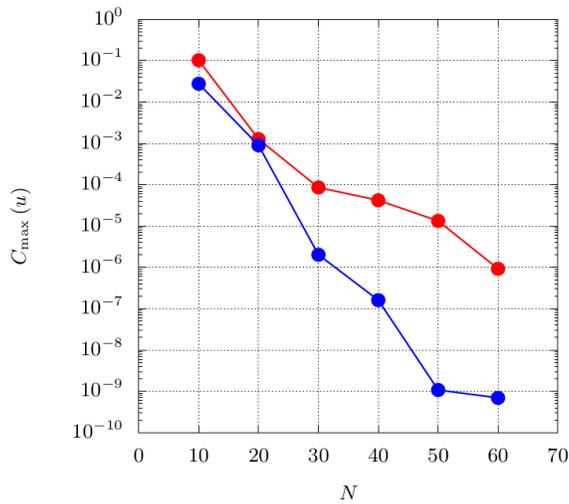}
\caption{Exponential decay of the maximum values of $C(u)$ for the affine-null code (blue) and the
Bondi code (red). Here we have set $\epsilon = 0.5$ for the initial data (\ref{eq:id_1}).}
\end{figure}

Another important feature we have verified is the conservation of the Newman-Penrose quantity
$c_{NP}$ (cf. (\ref{eq:NP})) for both the subcritical and supercritical solutions, using
the initial data (\ref{eq:id_2} ) with $\epsilon=2.273172$
(subcritical) and $\epsilon=2.273250$ (supercritical). In both cases, the
log-linear plots of the relative error 
$$\delta c_{NP}(u) =|c_{NP}(0)-c_{NP}(u)|/c_{NP}(0)
$$
shown in Fig's. 3(a) and
3(b) confirm that $c_{NP}=\mathrm{constant}$.
For the supercritical solution the final Bondi mass is
$\approx 0.0137$.  After black hole
formation the asymptotic quantity $c_{NP}$ is is not defined. In both simulations, we have used the
domain decomposition algorithm with $N_1=N_2=200$, $\lambda_0=L_0=1.0$.

\begin{figure}
\includegraphics[width=8cm,height=6cm]{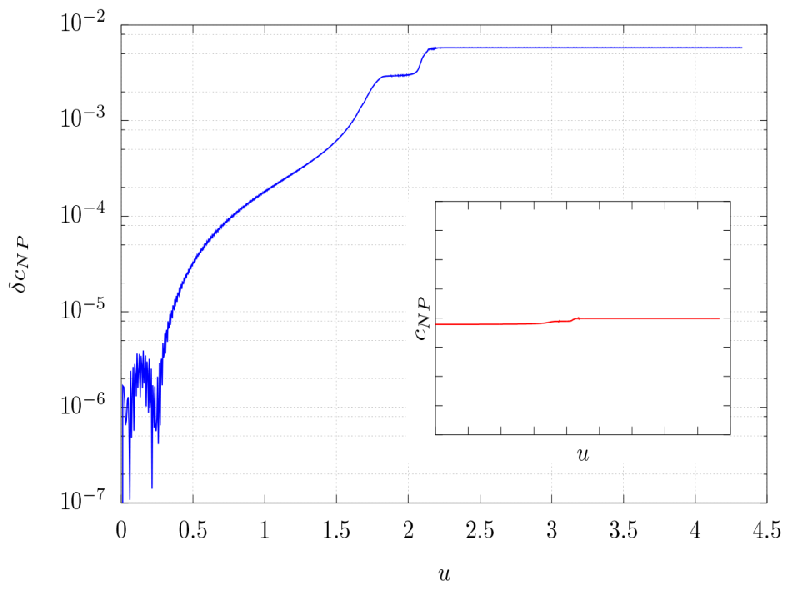}
\includegraphics[width=8cm,height=6cm]{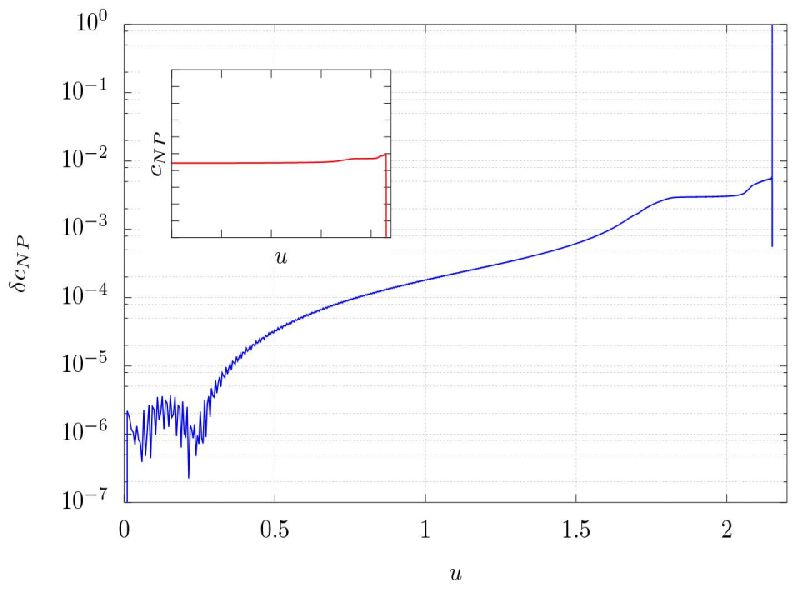}
\caption{The panels on the left and right, respectively, show the evolution of the relative variation $\delta c_{NP}$
of the Newman-Penrose quantity for subcritical and supercritical solutions. In the corresponding insets, 
the conservation of $c_{NP}$ is manifest. After black hole formation in the right panel, $c_{NP}$ is ill defined.
These solutions were generated using $\epsilon=2.273172, 2.273250$ in the initial data (\ref{eq:id_2}).}
\end{figure}

We identify the formation of a black hole in supercritical solutions by monitoring the limit  $H(u) \rightarrow 0$ as
$u \rightarrow u_E$ on approach to
the event horizon. In terms of the global behavior
of the metric function $r(u,\lambda)$, the asymptotic function $H(u)$ is computed in terms of the coefficients
$b_k^{(2)}(u)$ by evaluating 

\begin{equation}
H(u) = \lim_{\lambda \rightarrow \infty}\,\frac{r^{(2)}(u,\lambda)}{\lambda}.
\end{equation}

\begin{figure}
\includegraphics[width=8.5cm,height=6.5cm]{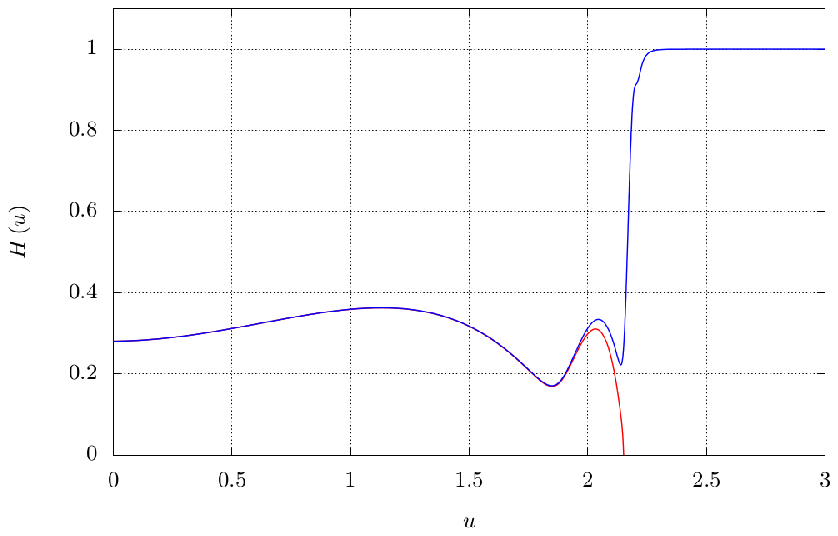}
\caption{Evolution of $H(u)$ for the subcritical (blue) and supercritical (red) solutions of Fig. 3.
For the subcritical case, $H(u) \rightarrow 1$ as the scalar field disperses and $r \rightarrow \lambda$.
For the supercritical case $H\rightarrow 0$ signaling the
infinite red shift as the event horizon forms.}
\end{figure}

\noindent As an illustration, we graph $H(u)$ in Fig. 4 for the subcritical (blue) and supercritical (red) solutions
considered in Fig. 3. Note that due to the closeness of the initial subcritical and supercritical amplitudes,
both curves almost coincide until $H(u)\rightarrow 0$ abruptly as the event horizon forms in the supercritical case.
For the subcritical case, $r \rightarrow \lambda$
and $H(u) \rightarrow 1$ as the scalar field disperses. This rapid divergence in the behavior of $H(u)$
for these two cases is expected from the instability associated with the attractor underlying critical collapse.

\begin{figure}
\includegraphics[width=8.5cm,height=6.5cm]{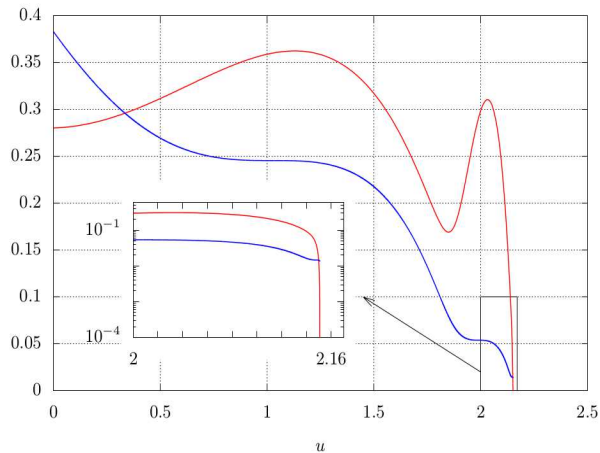}
\caption{The Bondi mass $M_B(u)$ (blue) and  $H(u)$ (red) for the slightly supercritical solution of Fig. 3.
The inset, which zooms into the interval just before horizon formation, shows that the Bondi mass approaches a
small but non-zero value as $H(u)$ approaches zero.}
\end{figure}

For the supercritical solutions, the behavior of $H(u)$ provides a criterion to determine
the final Bondi mass of the black hole. Recalling that $H(u)$ is positive and approaches
zero as the horizon forms, we can numerically determine the moment when $H(u)$ reaches
its smallest value and compute the corresponding value of the Bondi mass. In Fig. 5 we
depict a typical decay of the Bondi mass together with $H(u)$ for the slightly
supercritical solution generated with $\epsilon=2.273250$ in the initial data (\ref{eq:id_2}).
The Bondi mass decreases and reaches $M_B \approx 0.01375$. The inset shows
the log-linear plot of the interval near the formation of the horizon where $H(u)$ decays rapidly to
$\approx 10^{-8}$ and the Bondi mass approaches its small but non-zero final value. 
\begin{figure}[htb]
\includegraphics[width=8cm,height=7cm]{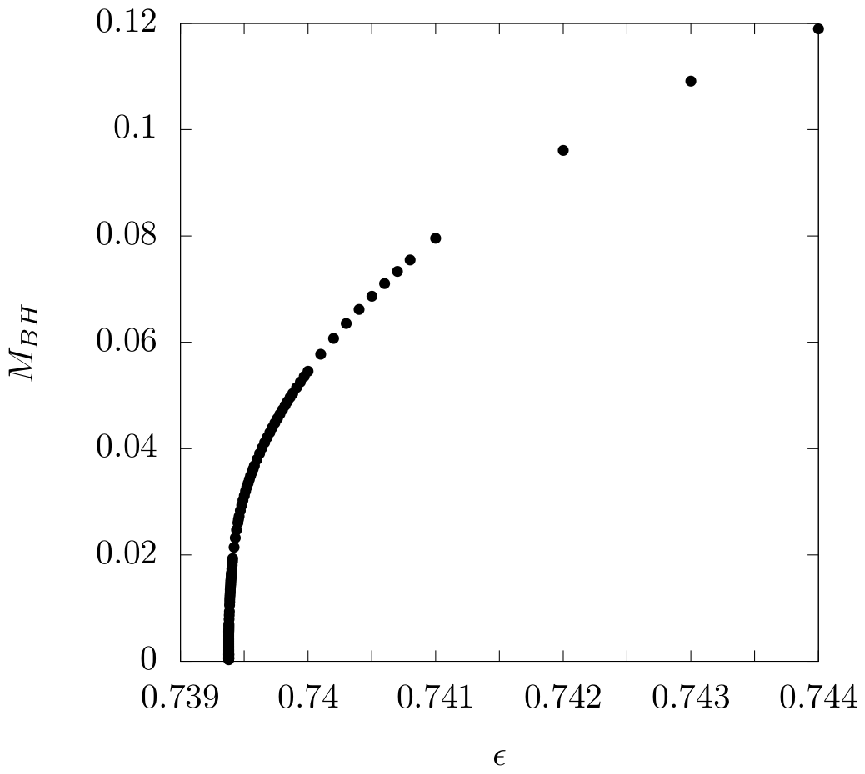}
\caption{Behavior of the final Bondi mass of the
black hole as a
function of the initial amplitude $\epsilon$ for the initial data  (\ref{eq:id_1}).}
\end{figure}

In order to reproduce the key features of critical
collapse, we have varied the initial amplitude $\epsilon$ to select the supercritical solution
with the smallest final Bondi mass $M_{BH}$ of the black hole using the above criterion.
In this process, we find an approximate value of the critical amplitude
$\epsilon_* \approx 0.7393775894$ for the initial data (\ref{eq:id_1})
using $N_1=N_2=150$ collocation points in each domain. The numerical results
plotted in Fig. 6 give an overall view of the results for $M_{BH}$.

The main feature of critical collapse is the Choptuik scaling law which relates the
mass $M_{AH}$ of the apparent horizon to the critical parameter according
to $M_{AH}=\bar{\kappa} (\delta \epsilon)^\gamma$, where
$\delta \epsilon=\epsilon-\epsilon_*$, $\bar{\kappa}$ is a constant depending upon
the initial data and $\gamma$ is the critical exponent.
This scaling law is also reflected in the final Bondi mass, as more clearly viewed in
the log-log plot of Fig. 7(a) constructed with the numerical data of Fig. 6.
The figure shows excellent agreement with the scaling law until $\delta \epsilon$
becomes very small on approach to the critical solution and the final Bondi mass
cannot be accurately resolved. Furthermore,
Gundlach \cite{gundlach2}, and Hod and Piran \cite{hod_piran} have predicted that
the original Choptuik scaling law is modified due to the discrete self-similarity (DSS)
nature of the type II critical solution. They have  proposed the following scaling law
\begin{equation}
\ln(M_{AH}) = \gamma \ln(\delta \epsilon) + f(\delta \epsilon) + \kappa
\end{equation}
\noindent where $\kappa$ is a constant that depends on the initial data family,
$f$ is a oscillatory function with period $\varpi = \Delta/2 \gamma$, and $\Delta$ is
the echoing period of the DSS critical solution. Hod and Piran \cite{hod_piran} have
verified this scaling law numerically and obtained
$\gamma \approx 0.37$, $\varpi \approx 4.61$ and $\Delta \approx 3.44$. Later 
P{\" u}rrer et al.~\cite{purrer} showed that this scaling law also closely
applies to the final Bondi mass $M_{BH}$ in the asymptotically flat treatment of critical collapse.
They argue that this result holds because the final stage of critical collapse is dominated
by the small region inside the DSS horizon.

We were able to identify the superposed oscillatory component in the numerical data for $M_{BH}$
by subtracting out the
$\gamma \ln (\delta \epsilon)$ term in the scaling law. The result
is shown in Fig. 7(b). We obtain the critical exponent $\gamma \approx 0.37134$, the period
of the oscillatory component $\varpi \approx 4.689$ and the echoing period $\Delta \approx 3.482$.
Our results differ by about
$1.7\%$ from those of P{\"u}rrer et al.~\cite{purrer}, which is accountable since we obtained
them with only 300 grid points, 150 in each subdomain,
while they used 10,000 points together with mesh refinement. Our results also show that a non-zero
Newman-Penrose constant does not effect universal critical behavior.

\begin{figure}
\includegraphics[width=7.8cm,height=7cm]{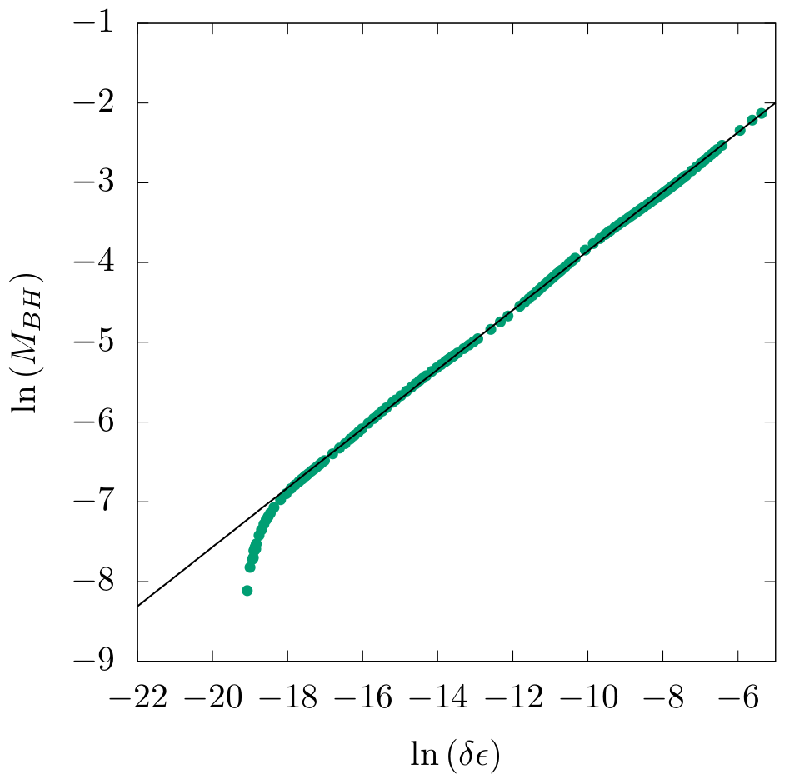}\includegraphics[width=8.5cm,height=7.3cm]{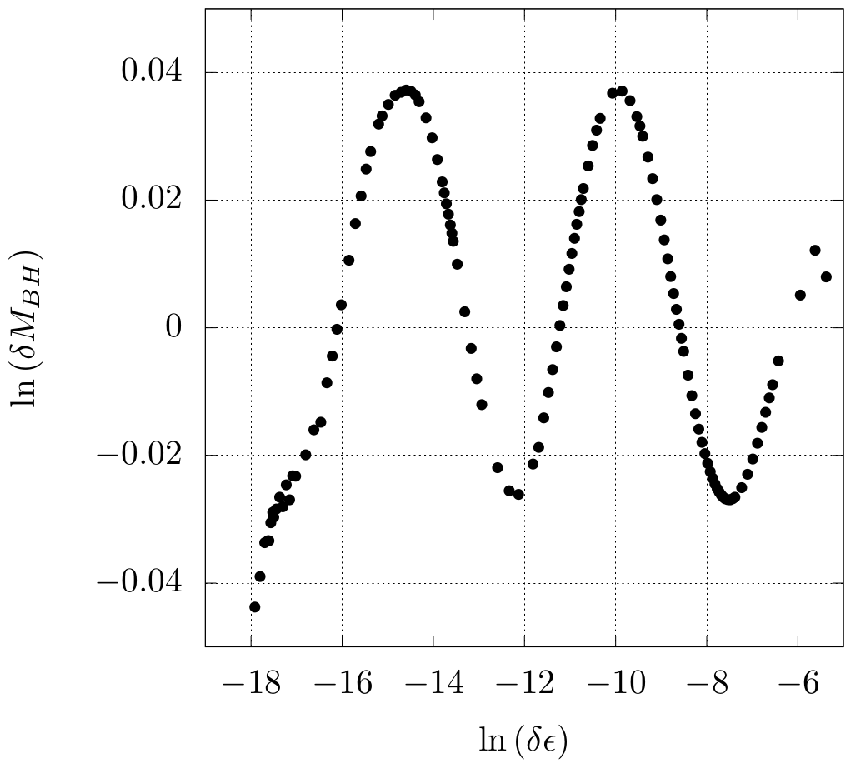}
\caption{Panel on the left: main scaling law $M_{BH} = \bar{\kappa}(\delta \epsilon)^\gamma$.
Panel on the right: oscillatory component $f(\delta \epsilon)$, where 
$\delta M_{BH} = M_{BH} - \bar{\kappa} (\delta \epsilon)^\gamma$.}
\end{figure}

\begin{figure}
\includegraphics[width=8cm,height=7cm]{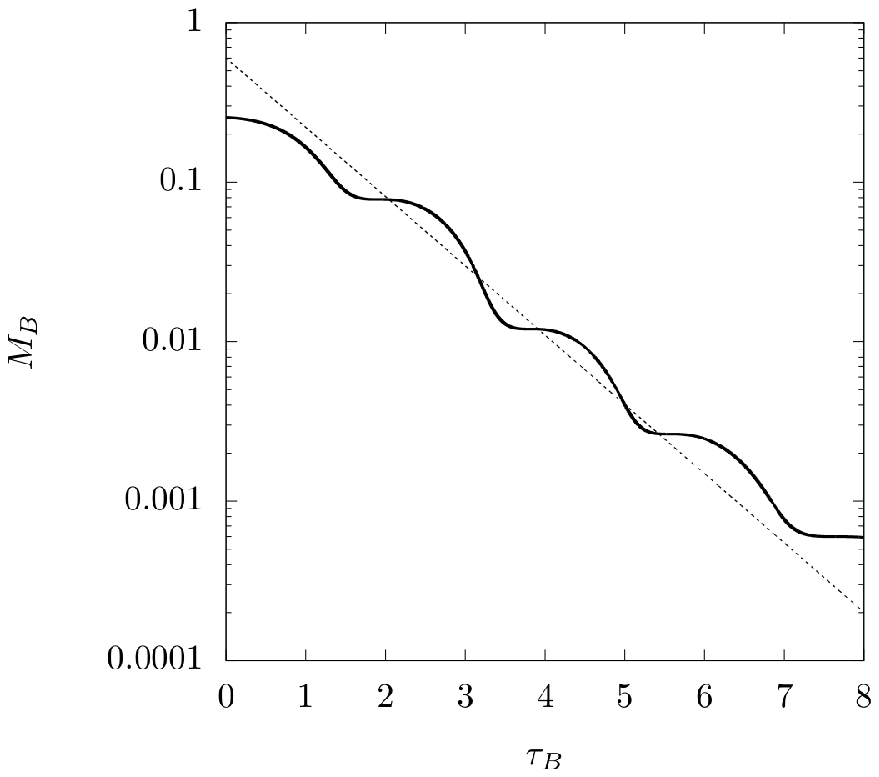}\includegraphics[width=8cm,height=7cm]{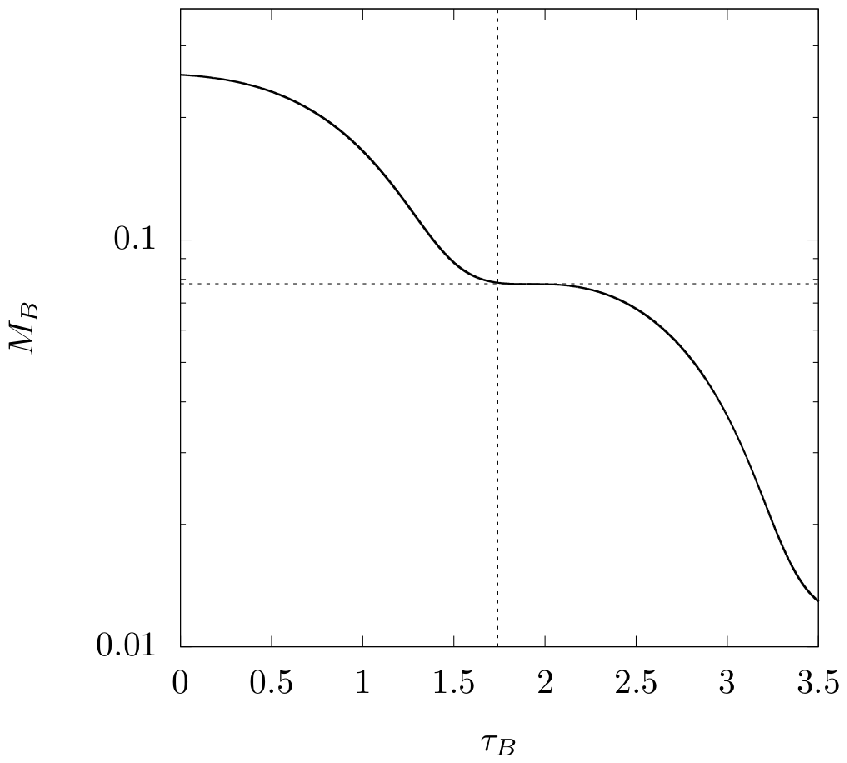}
\caption{Panel on the left: exponential decay of the Bondi mass represented by $M_B \propto \mathrm{e}^{-\tau_B}$.
Panel on the right: oscillation of the Bondi mass with respect to $\tau_B$ with an approximate period $\Delta/2$.
Here $\epsilon_*=0.7393775916$ for the initial data (\ref{eq:id_1}). }
\end{figure}

P{\" u}rrer et al.~\cite{purrer} found another aspect of the behavior of the critical solution when
described in terms of an adapted time coordinate
\begin{equation}
\tau_B = -\ln\left(\frac{u^*-u}{u}\right),
\end{equation} 
\noindent where $u^*$ is the accumulation time of DSS. 
They showed numerically that the Bondi mass decays exponentially in $\tau_B$,
together with an oscillatory component with period $\Delta/2$. We have confirmed this feature,
as illustrated in Fig. 8 for the decay of the Bondi mass in the near critical solution with
$\epsilon=0.7393775916$  for the initial data (\ref{eq:id_1}). In the left plot of Fig. 8, the dotted line describes
$M_B \propto \mathrm{e}^{-\tau_B}$ and the superposed oscillations have an approximate period of $\Delta/2$.
The right plot zooms in on the final approach to the
black hole.

We considered the formation of black holes using the initial data (\ref{eq:id_2}) with increased resolution 
by setting $N_1=N_2=200$ collocation points in each subdomain and setting the map parameter $L_0=0.15$.
We summarize the results in Fig. 9 by presenting the scaling law (left panel) and the oscillatory
component (right panel). The numerical parameters
are the critical exponent $\gamma \approx 0.3709$ and the oscillatory period $\varpi \approx 4.606$, resulting in an
echoing period $\Delta \approx 3.417$. All these parameters agree with the results 
in the work of P{\"u}rrer et al.~\cite{purrer}.

\begin{figure}[htb]
	\centering
	\includegraphics[width=8cm,height=7cm]{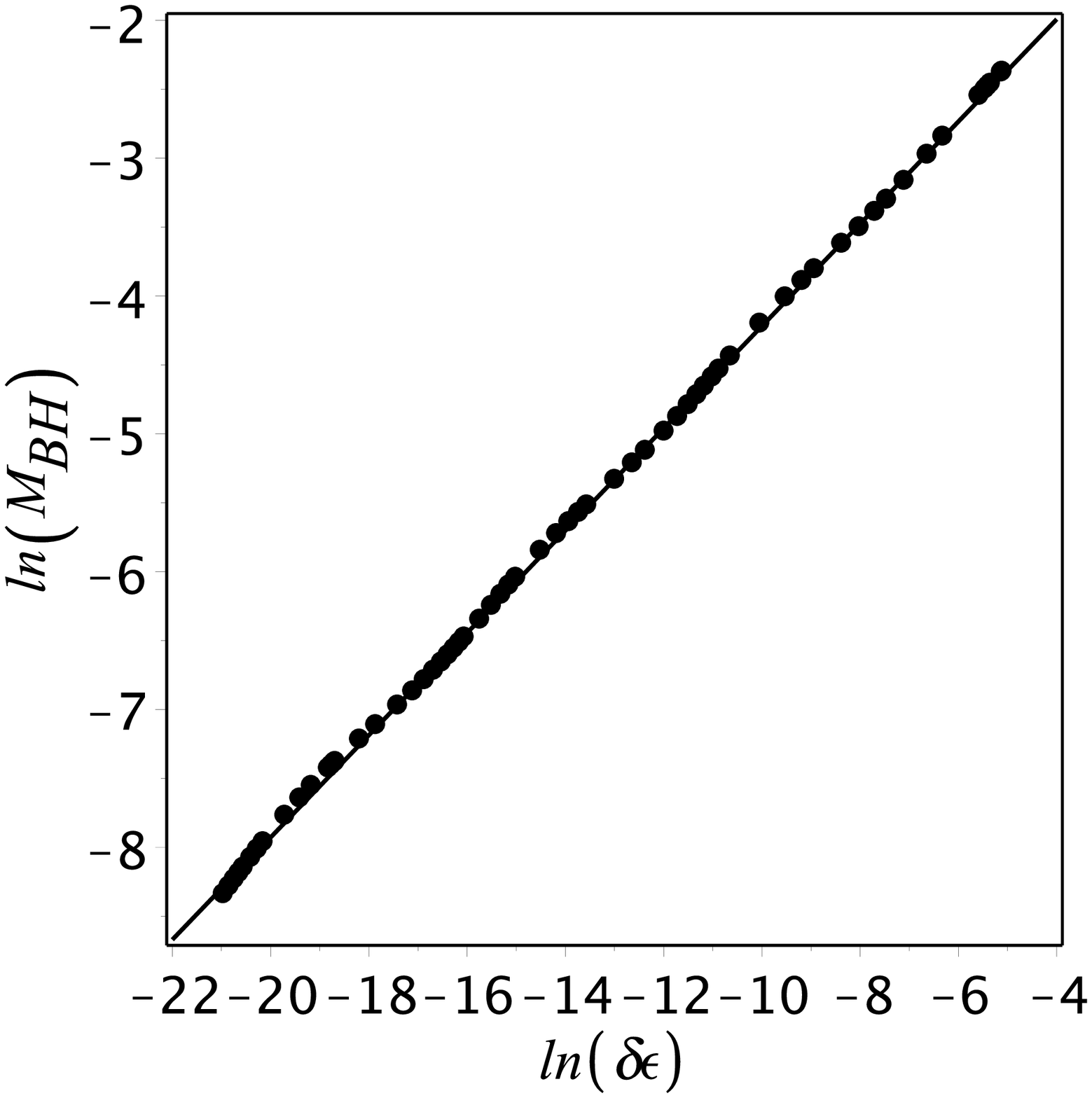}\includegraphics[width=8cm,height=7cm]{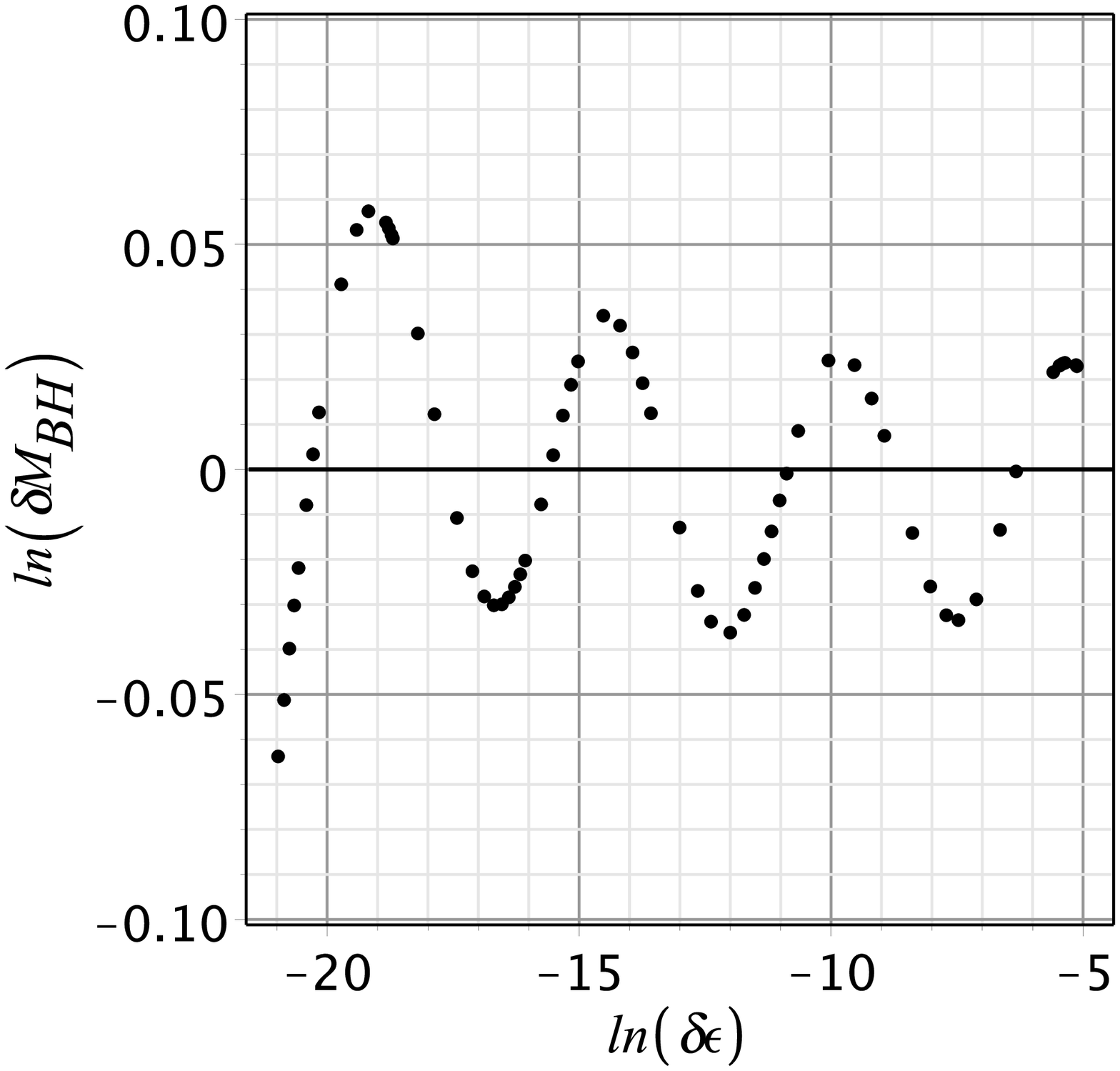}	
	\caption{Panel on the left: scaling law $M_{BH}=\bar{\kappa}(\delta \epsilon)^\gamma$,
	where $\gamma \approx 0.3709$ and $\epsilon_* \approx 2.2731656922$ is the approximate critical amplitude.
	Panel on the right: oscillatory component $f(\delta \epsilon)$.}
\end{figure}

We repeated this numerical experiment choosing initial data  (\ref{eq:id_3}) corresponding to $k=5$.
The approximate critical amplitude is $\epsilon_* \approx  0.458983605$
and $\gamma \approx 0.3706$. In this case, the initial
oscillations in the Chebyshev polynomials lead to a
large amount of ingoing radiation, some of which would cross the horizon in the supercritical case.
We present a graph of the scaling law for the final Bondi mass using 400 grid points in Fig. 10. 
Again, as in Fig. 7, there is excellent agreement with
the scaling law until $\delta \epsilon$ becomes very small on approach 
to the critical solution and the final Bondi mass cannot
be accurately resolved.

This result relates to an open issue raised by
P{\"u}rrer et al.~\cite{purrer}. They point out that simulations close to critical collapse
prior to their work confirmed
that black holes with arbitrarily small apparent horizons could be formed. But
that  left open tthe question whether
black holes with arbitrarily small Bondi mass could be formed. Their numerical simulations
for near critical collapse gave a small final
Bondi mass but did not resolve whether a Bondi mass gap might be necessary to correct
the scaling law. Numerically, this is a delicate issue since the exact critical solution harbors
is not known and, in the asymptotically flat context, might harbor a naked singularity.
They conjectured that radiation crossing the outer region of the
event horizon, outside the influence of the DDS behavior, might restrict the formation of black holes
with arbitrarily small Bondi mass. If that were the case then the transition between subcritical and
supercritical initial data would be discontinuous, i.e. it would be a transition
between subcritical dispersion with
zero final Bondi mass and supercritical collapse
to a black hole with
non-zero final Bondi mass. 

This leads to an interesting confluence between the analytical and
numerical results for the spherically symmetric collapse of a massless scalar field.
A theorem of Christodoulou
states that if the final Bondi mass is non-zero then
a black hole with regular event horizon forms~\cite{christo}.
Other analytic results of Christodoulou in the asymptotically flat context, establish
that naked singularities do occur in this problem, i.e.
the outgoing null cone from the central world line becomes singular,
although all prior outgoing null
cones extend non-singularly to ${\mathcal I}^+$,
with unbounded curvature as they approach the
singularity~\cite{christo3}. Christodoulou did not directly relate these results
to the Choptuik problem but they suggest
that the transition between the subcritical and
supercritical cases takes place through
this type of singular spacetime, as previously found in the numerical study
of the non-asymptotically flat, pure DSS
problem~\cite{gundlach2}. This scenario is consistent with the
global numerical study by Frolov and Pen~\cite{frolov}, although they do not
explicitly compute the Bondi mass. Our results for the final Bondi mass shown
in Fig. 10 show to high numerical accuracy that, even for initial data with
a large amount of ingoing radiation, there is no
Bondi mass gap in the transition between subcritical and supercritical
evolution. In this simulation, the initial Bondi mass is approximately $0.12108$ and falls between
three to four orders of magnitude in the near critical evolution.
\begin{figure}[htb]
	\centering
	\includegraphics[width=8cm,height=7cm]{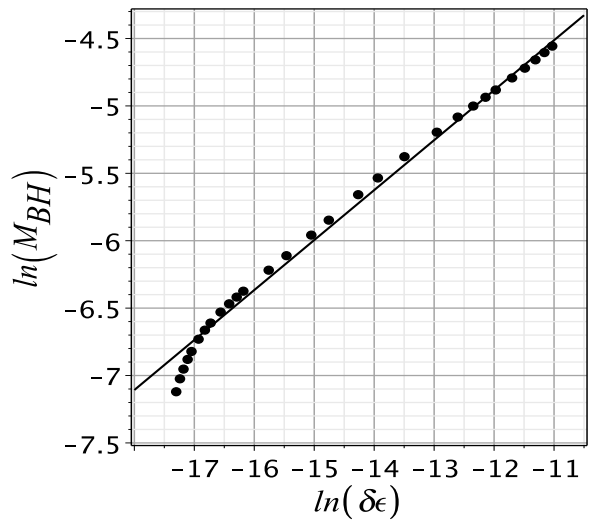}
	\caption{Scaling law for the final Bondi mass after evolving the
	initial data (\ref{eq:id_2}) with $k=5$. There is no evidence for a Bondi
	mass gap in the critical transition.}
\end{figure}

\section{Discussion}

In the context of spherically symmetric spacetimes with a massless scalar field,
we applied a new characteristic evolution algorithm based upon an affine parameter instead
of the areal coordinate of the Bondi-Sachs formulation. The advantages over
the Bondi-Sachs version were discussed. In particular, the hierarchical structure of the Bondi-Sachs
field equations is maintained by introducing variables which lead to unexpected quadratures
and a system of equations which are regular throughout the spacetime, up to the final singularity
in the case of gravitational collapse. Global regularity of the underlying equations
heuristically explains the vanishing of the final scalar monopole moment, which is a corner stone of the
no hair scenario for black holes. It allows a nonsingular treatment of the event horizon
and black hole formation, as opposed to the Bondi-Sachs system which degenerates on the
event horizon. In addition, the equations are simpler and are shown to lead to a more accurate
numerical treatment.

We implemented an innovative domain decomposition evolution algorithm based upon the
Galerkin-collocation method. After validating the code, we reproduced the main aspects of critical
collapse such as the mass scaling law and its oscillatory component  resulting from the discrete
self-similarity of the critical solution. The combination of the new set of the horizon penetrating
equations with spectral domain decomposition algorithm allows exhibiting these features in a 
grid with four hundred collocation points.
 
This allowed study of 
previously unexplored global features of the Choptuik problem for critical collapse of the scalar
field.  We showed that the effect of a non-vanishing Newman-Penrose constant does not affect
universal critical behavior. In addition, to high numerical accuracy, our results indicate, in the context of
an asymptotically flat exterior, that the final Bondi mass vanishes in the limit of
critical collapse, i.e. the critical case has no Bondi mass gap. This complements the
analogous result for measurements of the size of the apparent horizon for the Choptuik problem.
Our study encourages the application of the affine-null system to other problems

\begin{acknowledgments}

J. Crespo acknowledges the financial support of the Brazilian agency
Funda\c c\~ao Carlos Chagas Filho de Amparo \`a Pesquisa do Estado do Rio de Janeiro (FAPERJ).
H. P. de Oliveira thanks
Conselho Nacional de Desenvolvimento Cient\'ifico e Tecnol\'ogico (CNPq)
and Funda\c c\~ao Carlos Chagas Filho de Amparo \`a Pesquisa do Estado do Rio de Janeiro (FAPERJ)
(Grant No. E-26/202.998/518 2016 Bolsas de Bancada de Projetos (BBP)).
JW was supported by NSF grants PHY-1505965
and PHY-1806514 to the University of Pittsburgh. 
\end{acknowledgments}

\end{document}